\documentclass[a4paper,11pt]{article}

\usepackage{booktabs} 
\usepackage{colortbl} 
\usepackage[table]{xcolor} 
\usepackage{xfrac}
\usepackage[utf8]{inputenc}
\usepackage{amsmath,amsthm, amssymb, amsfonts, amsbsy}
\usepackage[mathscr]{euscript}
\usepackage{xspace}
\usepackage{graphicx,subfigure}
\usepackage{color}
\usepackage{enumerate}
\usepackage[shortlabels]{enumitem}
\usepackage[english]{babel}
\usepackage{cite}
\usepackage{verbatim}
\usepackage[margin=1.2in]{geometry}
\usepackage{authblk}
\usepackage{hyperref}
\usepackage{float}
\usepackage{soul}
\usepackage{rotating}
\usepackage{caption}
\usepackage[draft]{fixme}
\usepackage{longtable}
\usepackage{mathtools}

\fxsetup{
  layout=marginnote,
  marginface=\normalfont\tiny,
  envface=,
  inlineface=,
  innerlayout=noinline,
}

\definecolor{lightgray}{gray}{0.9}
\restylefloat{table}

\usepackage{algorithm}
\usepackage[noend]{algpseudocode}

\newcommand{\R}{\mathbb{R}}
\newcommand{\Z}{\mathbb{Z}}

\newcommand{\Q}{\mathbb{Q}}

\newcommand{\NP}{\mathscr{NP}}
\newcommand{\Pclass}{\mathscr{P}}
\newcommand{\bigO}{\mathcal{O}}

 \let\mathscr\relax
\usepackage[scr]{rsfso}

\newcommand{\minmaxBCPk}{\textsc{min-max BCP}_k}
\newcommand{\minmaxBCPtwo}{\textsc{min-max BCP}_2}
\newcommand{\maxminBCPtwo}{\textsc{max-min BCP}_2}

\newcommand{\BCPk}{\textsc{BCP}_k}

\newcommand{\maxminBCPk}{\textsc{max-min BCP}_k}
\newcommand{\maxminBCP}[1]{\textsc{max-min BCP}_{#1}}
\newcommand{\minmaxBCP}[1]{\textsc{min-max BCP}_{#1}}
\newcommand{\maxminBCPkUnw}{\textsc{1-max-min BCP}_k}
\newcommand{\minmaxBCPkUnw}{\textsc{1-min-max BCP}_k}

\newcommand{\minmaxBCPUnw}[1]{\textsc{1-min-max BCP}_{#1}}

\newcommand{\minmaxBCPsemk}{\textsc{min-max BCP}}
\newcommand{\maxminBCPsemk}{\textsc{max-min BCP}}

\newcommand{\eps}{{\varepsilon}}
\newcommand{\floor}[1]{{\left\lfloor #1 \right\rfloor}}
\newcommand{\ceil}[1]{{\left\lceil #1 \right\rceil}}
\newcommand{\bea}{\begin{eqnarray*}} 
\newcommand{\eea}{\end{eqnarray*}}
\newcommand{\bean}{\begin{eqnarray}}  
\newcommand{\eean}{\end{eqnarray}}
\newcommand{\beq}{\begin{equation}} 
\newcommand{\eeq}{\end{equation}}

\newcommand{\cala}{{\cal A}}

\newcommand{\calc}{{\cal C}}

\newcommand{\calg}{{\cal G}}

\newcommand{\calp}{{\cal P}}

\newcommand{\calae}{\ensuremath{\mathcal{A}_{\eps}}}

\providecommand{\keywords}[1]{\textit{Keywords:} #1}

\DeclareMathOperator{\opt}{\rm{OPT}}

\theoremstyle{plain}
\newtheorem{theorem}{Theorem}

\newtheorem{corollary}[theorem]{Corollary}
\newtheorem{lemma}[theorem]{Lemma}

\newtheorem{fact}{Fact}

\theoremstyle{definition}
\newtheorem{problem}{Problem}

\usepackage{silence}
\begin{document}

\title{Approximation and parameterized algorithms to find balanced
  connected partitions of graphs\protect\footnote{Research partially supported by grant \#2015/11937-9, S\~ao Paulo Research Foundation (FAPESP).}}


\author[1]{Phablo F. S. Moura\thanks{(phablo@dcc.ufmg.br).}}
\author[2]{Matheus J. Ota\thanks{(mjota@uwaterloo.ca)}}
\author[3]{Yoshiko Wakabayashi\thanks{(yw@ime.usp.br) supported by CNPq (Proc. 306464/2016-0 and 423833/2018-9).}} 
\affil[1]{Departamento de Ciência da Computação, Universidade Federal de Minas Gerais, Brazil}
\affil[2]{Department of Combinatorics and Optimization, University of Waterloo, Canada}
\affil[3]{Instituto de Matemática e Estatística, Universidade de São Paulo, Brazil}

\maketitle

\begin{abstract}
  Partitioning a connected graph into $k$~vertex-disjoint connected subgraphs of
  similar (or given) orders is a classical problem that has been intensively investigated since late seventies.  
  Given a connected graph~$G=(V,E)$ and a weight
  function~$w \colon V \to \Q_\geq$, a connected $k$-partition of~$G$ is
  a partition of~$V$ such that each class induces a connected
  subgraph.  The balanced connected $k$-partition problem consists in
  finding a connected $k$-partition in which every class has roughly
  the same weight.  To model this concept of balance, one may seek
  connected \hbox{$k$-partitions} that either maximize the weight of a
  lightest class $(\maxminBCPk)$ or minimize the weight of a heaviest
  class $(\minmaxBCPk)$.  Such problems are equivalent when~$k=2$, but
  they are different when~$k\geq 3$.  In this work, we propose a
  simple pseudo-polynomial $\frac{k}{2}$-approximation algorithm
  for~$\minmaxBCPk$ which runs in time~$\bigO(W|V||E|)$,
  where~$W = \sum_{v \in V} w(v)$.  Based on this algorithm and using
  a scaling technique, we design a (polynomial)
  $(\frac{k}{2} +\eps)$-approximation for the same problem with
  running-time~$\bigO(|V|^3|E|/\eps)$, for any fixed~$\eps>0$.
  Additionally, we propose a fixed-parameter tractable algorithm based
  on integer linear programming for the unweighted $\maxminBCPk$
  parameterized by the size of a vertex cover.
\end{abstract}


\section{Introduction}
The problem of partitioning a connected graph into a given number~$k\geq 2$ of connected subgraphs with prescribed orders
was first  studied by Lovász~\cite{Lovasz77} and Gy\"ori~\cite{Gyori78} in the late seventies.  
Let $[k]$ denote the set $\{1, 2, \ldots, k\}$, for every integer~$k \geq 1$. A \textit{connected $k$-partition} of a connected graph~$G=(V,E)$ is a partition of~$V$ into classes~$\{V_i\}_{i=1}^k$ of nonempty
subsets such that, for each~$i \in [k]$, the subgraph~$G[V_i]$ is connected, where~$G[V_i]$ denotes the subgraph of~$G$ induced by the vertices~$V_i$.

Consider a pair $(G,w)$, where $G=(V,E)$ is a connected graph and $w \colon V \to \Q_>$ is a function that assigns positive weights
to the vertices of~$G$. 
For each~$V' \subseteq V$, we define~$w(V')=\sum_{v \in V'} w(v)$. 
Furthermore, if~$G'=(V^\prime,E^\prime)$ is a subgraph of $G$, we write~$w(G')$ instead of~$w(V^\prime)$. 
If ${\calp} = \{V_i\}_{i \in [k]}$ is a connected $k$-partition of $G$, then
$w^+({\calp})$ stands for $\max_{i \in [k]} \left\{w(V_i)\right\}$, and
$w^-({\calp})$ stands for $\min_{i \in [k]} \left\{w(V_i)\right\}$.

The concept of balance of the classes of a connected partition of a graph can be expressed in different ways. 
In this work, we consider two related variants whose objective functions express this concept.

\begin{problem} \textsc{Min-Max Balanced Connected $k$-Partition} \rm{($\minmaxBCPk$)}\\
  \textsc{Instance:} a connected graph $G=(V,E)$, and a weight function $w \colon V \to \Q_{\geq}$. \\
  \textsc{Find:} a connected $k$-partition $\calp$ of $G$.\\
  \textsc{Goal:} minimize $w^+(\calp)$.
\end{problem}

\begin{problem} \textsc{Max-Min Balanced Connected $k$-Partition} \rm{($\maxminBCPk$)}\\
  \textsc{Instance:} a connected graph $G=(V,E)$, and a weight function $w \colon V \to \Q_{\geq}$. \\
  \textsc{Find:} a connected $k$-partition $\calp$ of $G$.\\
  \textsc{Goal:} maximize $w^-(\calp)$.
\end{problem}




We remark that~$\minmaxBCPtwo$ and $\maxminBCPtwo$ are equivalent, that is, for any instance, an optimal solution for the min-max version is also an optimal solution for the max-min version.
If $k>2$ the corresponding optimal $k$-partitions may differ (the reader is referred to the examples given by Lucertini, Perl, and Simeone~\cite{LucPerSim93}).

Throughout this paper we assume that~$k \geq 2$. When $k$ is in the name of the problem, we are
considering that $k$ is fixed. 
The problems in which $k$ is part of the instance are denoted similarly but without
specifying $k$ in the name (e.g.~$\maxminBCPsemk$).

The unweighted (or cardinality) versions of these problems refer to
the case in which all vertices have equal weight, which may be assumed to be~$1$. 
We denote the corresponding problems as~$\minmaxBCPkUnw$ and~$\maxminBCPkUnw$.






In this paper, we show approximation algorithms for $\minmaxBCPk$ and
also mention approximation results for both $\minmaxBCPk$ and
$\maxminBCPk$. To make clear what we mean by an
\emph{$\alpha$-approximation} algorithm, we define this concept. We
also define another closely related concept that will be used here.

Let $\cala$ be an algorithm for an optimization problem $\Pi$. If $I$
is an instance for $\Pi$, we denote by $\cala(I)$ the value of the
solution produced by~$\cala$ for $I$, and by~$\opt(I)$ the value of an
optimal solution for~$I$.
If $\Pi$ is a minimization (resp. maximization) problem, and $\cala$
is a polynomial-time algorithm, we say that~$\cala$ is an
\emph{$\alpha$-approximation algorithm}, for some $\alpha\ge 1$, if
$\cala(I) \le \alpha \opt(I)$ (resp.  $\cala(I) \ge 1/\alpha \opt(I)$)
for every instance~$I$ of~$\Pi$.  We also say that $\cala$ is an
approximation algorithm with \emph{ratio} $\alpha$.  The approximation
ratio $\alpha$ need not be a constant: it can be a function
$\alpha(I)$ that depends on~$I$. So, whenever we refer to an
approximation algorithm, we mean that it runs in polynomial time on
the size of the instance.

When an approximation ratio $\alpha$ can be guaranteed for an
algorithm, but it may run in pseudo-polynomial time, we refer to it as
a \emph{pseudo-polynomial $\alpha$-approximation}. This is not a usual
terminology, but it will be appropriate for our purposes.

Problems of finding balanced connected partitions can be used to model a rich collection of applications in
logistics, image processing, data base, operating systems, cluster analysis, education, robotics and biological
networks~\cite{BecPer83,LucPerSim89,LucPerSim93,MarSimNal97,MatBoz12,ZhoWanDinHuSha19,MatGrb20}.

\subsection{Some  known results}

  We first mention some hardness results to solve or to obtain certain
  approximate solutions for the $k$-connected partition problems.
  Dyer and Frieze~\cite{DyeFri85} proved that~$\maxminBCPkUnw$ is
  $\NP$-hard on bipartite graphs.  Furthermore,~$\maxminBCPkUnw$ have
  been shown by Chleb\'\i kov\'a~\cite{Chl96} to be $\NP$-hard to
  approximate within an absolute error guarantee
  of~$n^{1- \varepsilon}$, for all $\varepsilon > 0$.  For the
  weighted versions, Becker, Lari, Lucertini and
  Simeone~\cite{BeckerLLS98} proved that~$\maxminBCP{2}$ is $\NP$-hard
  on grid graphs.  Wu~\cite{Wu12} showed that $\maxminBCPk$ is
  $\NP$-hard on interval graphs for every $k$. Chataigner, Salgado,
  and Wakabayashi~\cite{ChaSalWak07} showed that~$\maxminBCPk$ is
  strongly $\NP$-hard, even on $k$-connected graphs. Hence, unless
  $\Pclass = \NP$,~$\maxminBCPk$ does not admit a fully
  polynomial-time approximation scheme (FPTAS). They also showed that
  when $k$ is part of the instance, $\minmaxBCPsemk$ cannot be
  approximated within a ratio better than~$6/5$.


  Now, we turn to existential or algorithmic results for the
  unweighted versions of the $k$-connected partition problems.  When
  the input graph~$G$ is~$k$-connected, Gy{\"o}ri~\cite{Gyori78}, and
  Lov{\'a}sz~\cite{Lovasz77} proved that one can always find a
  connected~$k$-partition where each class has a prescribed number of
  vertices. It is not difficult to devise a polynomial-time algorithm
  to find a $2$-connected partition in which the classes have
  prescribed sizes. However, the proof of the existence of the 
  desired~$k$-connected partition on any~$k$-connected graph given by 
  Gy{\"o}ri~\cite{Gyori78} does not seem to yield a polynomial-time 
  algorithm.  (In the past, an algorithm for this problem was claimed 
  to be polynomial, but such result has not been established yet.)

  Polynomial-time algorithms have been designed for restricted cases
  of the result of Gy{\"o}ri~and Lov{\'a}sz. 
  Suzuki, Takahashi and Nishizeki~\cite{SUZUKI1990227} devised
  a linear-time algorithm to find a connected~$2$-partition on 
  a $2$-connected graph. When the input graph is~$3$-connected,
  Suzuki et al.~\cite{suzuki1990algorithm} presented a 
  quadratic-time algorithm to compute a connected~$3$-partition.
  If~$G$ is planar and~$4$-connected, 
  Nakano, Rahman and Nishizeki~\cite{NakRahNis97} shows that a
  connected $4$-partition can be found in linear-time.

  More recently, Chen et al.~\cite{chen2019approximation} designed
  a~$3/2$-approximation for~$\minmaxBCPUnw{3}$
  and~$24/13$-approximation for~$\minmaxBCPUnw{4}$. For
  $\minmaxBCPUnw{k}$, $k \geq 4$, they also provided a
  $k/2$-approximation.


  Now, let us consider the weighted versions.  For $\maxminBCPk$
  (resp. $\minmaxBCPk$), Perl and Schach~\cite{PerlS81} (resp.
  Becker, Schach, and Perl~\cite{BeckerPS82}) designed polynomial-time
  algorithm when the input graph is a tree. Also for trees,
  Frederickson~\cite{frederickson91} proposed linear-time algorithms
  for both~$\maxminBCPk$ and~$\minmaxBCPk$. Polynomial-time algorithms
  were also derived for $\maxminBCP{2}$ on graphs with at most two
  cut-vertices~\cite{Chl96, AliCal99}. For $\maxminBCPk$ on ladders, a
  polynomial-time algorithm was obtained by Becker et al. (2001).

  In 1996, Chlebíková~\cite{Chl96} designed a $4/3$-approximation
  algorithm for $\maxminBCP{2}$. More recently, Chen et
  al.~\cite{Chen20} have shown that the algorithm obtained by
  Chlebíková has approximation ratio $5/4$ for~$\minmaxBCP{2}$.
  These authors also obtained approximation algorithms with
  ratio~$3/2$ and $5/3$ for $\minmaxBCP{3}$ and $\maxminBCP{3}$,
  respectively. In 2012, Wu~\cite{Wu12} designed a FPTAS
  for~$\maxminBCP{2}$ restricted to interval graphs.

  In Table~\ref{table:approximation_table}, we present some of the
  results we have mentioned.  Other approximation algorithms have also
  been obtained, but they have slightly weaker ratios, or impose
  conditions on the input graph.

  When $k$ is part of the input, approximation algorithms were
  proposed by Bornd{\"o}rfer, Elijazyfer and
  Schwartz~\cite{BorEliSch19}. For both~$\maxminBCPsemk$
  and~$\minmaxBCPsemk$, their algorithms has approximation
  ratio~$\Delta$, where~$\Delta$ is the maximum degree of an arbitrary
  spanning tree of the input graph~$G$. Specifically
  for~$\maxminBCPsemk$, their $\Delta$-approximation only holds for
  instances in which the largest weight is at most $w(G)/(\Delta\,k)$.



\begin{table}
\centering
\caption{Approximation results for~$\BCPk$.}
\rowcolors{1}{lightgray}{}
\begin{tabular}{ c  c  c }
\hiderowcolors
\toprule
Author(s) & Approximation Ratio & Problem Version \\
\midrule
\noalign{\global\rownum=1}
\showrowcolors
Chen et al.~\cite{chen2019approximation} & $k/2$ & $\minmaxBCPUnw{k},~k \geq 3$\\
Chen et al.~\cite{chen2019approximation} & $24/13$ & $\minmaxBCPUnw{4}$\\
Chlebíková~\cite{Chl96} & $5/4$ & $\minmaxBCP{2}$\\
Chen et al.~\cite{Chen20} & $3/2$ & $\minmaxBCP{3}$\\
Chlebíková~\cite{Chl96} & $4/3$ & $\maxminBCP{2}$\\
Chen et al.~\cite{Chen20} & $5/3$ & $\maxminBCP{3}$\\
\bottomrule
\end{tabular}
\label{table:approximation_table}
\end{table}

The results we mentioned above were mostly concerned with
polynomial-time algorithms, approximation algorithms and hardness
results for the two variants of balanced connected $k$-partition problem. 

For completeness, we mention that some mixed integer linear
programming formulations were also proposed for these
problems (see~\cite{Mat14,ZhoWanDinHuSha19,MiyazawaMOW20,MiyazawaMOW21}).



 

\subsection{Our contributions}

We generalize the $k/2$-approximation algorithm for
$\minmaxBCPUnw{k}$, $k \geq 3$,  designed by Chen et
al.~\cite{chen2019approximation}, and present an approximation
algorithm for (the weighted version) $\minmaxBCP{k}$, $k \geq 3$. We
prove that it has basically the same approximation ratio: namely,
$k/2+\varepsilon$, for any arbitrarily small~$\varepsilon>0$. For
that, we use a scaling technique to deal with weights that might be
very large. When the weights assigned to the vertices of the input
graph are bounded by a polynomial on the order of the graph, it
achieves the ratio $k/2$.  A $3/2$-approximation algorithm for
$\minmaxBCP{3}$ was obtained by Chen et al.~\cite{Chen20}, but its
analysis and implementation are slightly more complicated than the
algorithm we show here.


We also prove that $\maxminBCPkUnw$ is fixed-parameter tractable when the parameter is the size of a vertex cover of the input graph.
This algorithm is based on integer linear programming and it has a doubly exponential dependency on the size of a vertex cover.
To the best of our knowledge, no other FPT algorithm for balanced connected partition problems is described in the literature.
Despite the fact that the proposed algorithm is not practical, the strategy used to model connected partitions may be applicable to show that other problems involving connectivity constraints are fixed-parameter tractable when the parameter is the size of a vertex cover.
\section{Approximation algorithm for $\minmaxBCPk$}


Chen et al.~\cite{chen2019approximation} devised an algorithm for
~$\minmaxBCPkUnw$ with approximation ratio~$\frac{k}{2}$.
This algorithm iteratively applies two simple operations, namely \textsc{Pull} and \textsc{Merge}, to reduce the size of the largest class.
In what follows, we show how to generalize such operations for the \emph{weighted} case to design a $(\frac{k}{2}+\varepsilon)$-approximation for $\minmaxBCPk$, for any $\varepsilon>0$.

First we discuss the algorithm for the case~$k = 3$,  and then we show how to use the con\-nec\-ted $3$-partition returned by this algorithm to obtain a connected $k$-partition for any~$k \geq 4$.

Throughout this section, $k$ is a positive integer, and $(G, w)$ denotes an instance of $\minmaxBCPk$, where $G=(V,E)$ and $w\colon V \to \Q_\geq$. Also, when convenient, we  denote by $W$ the sum of the weights of the vertices in $G$, that is, $W=w(G)$.
In this section, we assume without loss of generality that $w$ is an integer-valued function (otherwise, we  may simply multiply all weights by the least common multiple of the denominators). 
The following trivial fact is used to show the approximation ratio of the algorithms for $\minmaxBCPk$ proposed here. When convenient, we denote
by  $\opt_k(I)$ the value of an optimal solution for an instance $I$ of $\minmaxBCPk$.

\begin{fact}\label{fact:opt}
Any  optimal solution for an instance $I=(G,w)$ of $\minmaxBCPk$ has value at least~$w(G)/k$, that is, $\opt_k(I) \geq w(G)/k$.
\end{fact}

For $k\geq 3$, let ${\calg}_k$ be the class of connected graphs~$G$ containing a cut-vertex $v$ such that $G-v$ has at least $k-1$ components. 
We denote by $c(H)$ the number of components of a graph $H$. 
The next lemma provides a lower bound for the value of an optimal solution of $\minmaxBCPk$ on instances $(G,w)$ with $G\in{\calg}_k$.



\medskip

\begin{lemma}\label{lemma:bound:max}
  Let $I=(G,w)$ be an instance of $\minmaxBCPk$ in which $G\in{\calg}_k$, 
  and $v$ is a cut-vertex of $G$ such that $c(G-v) = \ell \geq k-1$. 
  Let~${\calc}=\{C_i\}_{i\in[\ell]}$ be the set of the components
  of $G-v$.  Suppose further that $w(C_i)\leq w(C_{i+1})$ for every
  $i \in [\ell-1]$. 
  Then every connected $k$-partition $\mathcal{P}$ of $G$ satisfies
\[w^+(\mathcal{P}) \geq w(v) + \sum_{i \in [\ell - k + 1]} w(C_i).\]
In particular, $\opt_k(I)\geq w(v) + \sum_{i \in [\ell - k + 1]} w(C_i).$
\end{lemma}

\begin{proof}
  Consider a connected $k$-partition~$\mathcal{P}$ of~$G$, and let $V^*$ be the class in $\mathcal{P}$  that contains~$v$. Let
      $q^* :=|\{C \in {\calc} \colon V(C) \subseteq V^*\}|$  and 
      $q:= |\{C \in \mathcal{C} \colon V(C) \nsubseteq V^*\}|$.

      Note that each class in $\mathcal{P}\setminus\{V^*\}$ is either a component of $G-v$ or it is a set properly contained in a unique component of $G-v$.  
      Hence, $q^* + q = \ell$ and
      $q \leq k - 1$.  
      Therefore, $q^* =\ell - q \geq \ell - k + 1$.
    Since $w(C_1)\leq w(C_2) \leq \ldots \leq w(C_\ell)$, we conclude that $w(V^*) \geq w(v) + \sum_{i \in [\ell - k + 1]} w(C_i)$, and therefore, $w^+(\mathcal{P}) \geq w(V^*)$.
    Clearly, it holds that~$\opt_k(I)\geq w(v) + \sum_{i \in [\ell - k + 1]} w(C_i)$.
\end{proof}

We now present an algorithm for $\minmaxBCP{3}$ that  generalizes the
algorithm proposed by Chen et al.~\cite{chen2019approximation} for the
unweighted version of this problem. We adopt basically the same notation used by
these authors to refer to the basic operations which are the core of
the algorithm.

The strategy used in the algorithm is to start with an arbitrary connected $3$-partition and
improve it by applying successively (while it is possible) the operations
\textsc{Merge} and \textsc{Pull}, defined in what follows.

We say that a connected 3-partition $\{V_1, V_2, V_3\}$ of $G$ is
\emph{ordered} if $w(V_1) \leq w(V_2) \leq w(V_3)$.  The input for
\textsc{Pull} and~\textsc{Merge} is an ordered connected 3-partition
$\{V_1, V_2, V_3\}$.  As these operations may be applied several
times, a reordering of the classes is performed at the end, if
necessary. In this context, we say that an ordered
3-partition~$\mathcal{P}=\{V_1, V_2, V_3\}$ is \emph{better} than an
ordered 3-partition~$\mathcal{Q}=\{X_1, X_2, X_3\}$ if
$w(V_3)< w(X_3)$.

We say that two classes $V_i$ and $V_j$ are adjacent if there is an
edge in $G$ joining these classes. For $X\subset V$, we denote by
$N(X)$ the set of vertices in $G$ that are adjacent to a vertex of
$X$.
\\

\begin{itemize}
    \item $\textsc{Merge}(\mathcal{P})$
      \begin{itemize}
        \item[--] {Input}: an ordered connected $3$-partition $\mathcal{P}=\{V_1, V_2, V_3\}$ of $G$.
        \item[--] {Pre-conditions}: (a)~$w(V_3) > w(G)/2$; (b)~$|V_3| \geq 2$; (c)~$V_1$ and $V_2$ are adjacent.
        \item[--] {Output}: a connected 3-partition~$\{V_1 \cup V_2, V_3', V_3^{\prime \prime}\}$, where~$\{V_3', V_3^{\prime \prime}\}$ is an arbitrary connected 2-partition of~$G[V_3]$. Reorder the classes if necessary, and return an ordered partition. 
    \end{itemize}
\end{itemize}
The $3$-partition returned by \textsc{Merge} is better than the input  partition since~$w(V_3') < w(V_3)$,~$w(V_3^{\prime \prime}) < w(V_3)$ and~$w(V_1) + w(V_2) < \frac{w(G)}{2} < w(V_3)$.
Note that a depth-first search suffices to check the pre-conditions.
Moreover, a connected~$2$-partition of~$G[V_3]$ can be easily obtained from any spanning tree of this graph.
Hence \textsc{Merge} can be executed in~$\bigO(|V| + |E|)$.


          
%
%
\begin{itemize}    
    \item $\textsc{Pull}(\mathcal{P}, U,i)$
      \begin{itemize}
        \item[--]{Input}: an ordered connected $3$-partition $\mathcal{P}=\{V_1, V_2, V_3\}$ of $G$, a nonempty subset $U$ of vertices,  and $i\in\{1,2\}$.
        \item[--] {Pre-conditions}: (a)~$w(V_3) > {w(G)}/{2}$; (b)~$U \subsetneq V_3$,  $G[V_i\cup U]$ and~$G[V_3 \setminus U]$ are connected; (c)~$w(V_i\cup U) < w(V_3)$.
        \item[--] {Output}: a connected 3-partition~$\{V_j, V_i \cup U, V_3 \setminus U\}$ where~$j\in \{1,2\}\setminus \{i\}$. Reorder the classes if necessary, and return an ordered partition.
          
    \end{itemize}
  \end{itemize}
  Note that $\textsc{Pull}(\mathcal{P}, U,i)$ improves the input
  partition $\mathcal{P}=\{V_1, V_2, V_3\}$, since~$w(V_3 \setminus U) < w(V_3)$,
  $w(V_j) < w(V_3)$ and $w(V_i\cup U) < w(V_3)$. Moreover, it is only executed when
  a set $U$ satisfying the pre-conditions is given. Thus, this
  operation can be executed in $\bigO(|V|)$ time.  We next
  discuss the time complexity to find such a set~$U \subsetneq V_3$, if it exists.
  For simplicity, we say that such a set~$U$  is~\emph{pull-admissible} (w.r.t.~$i$).
  Observe that if~$V_3$ contains a pull-admissible subset, then~$V_3$ has at least 2 vertices.




\begin{algorithm}[H]
\hspace*{\algorithmicindent} \textbf{Input:} An ordered connected 3-partition~$\mathcal{P} = \{V_1, V_2, V_3\}$ of $(G,w)$, and $i\in\{1,2\}$.\\
\hspace*{\algorithmicindent}
\textbf{Output:} Either a set $U \subset V_3$ that is pull-admissible w.r.t.~$i$, or the emptyset  $\emptyset$.

\begin{algorithmic}[1]
\Procedure {\textsc{PullCheck}}{$\mathcal{P},i$}

\For {$v \in N(V_i) \cap V_3$}  \label{line:for}
	\State {Let~$\mathcal{C} = \{C_1, \ldots, C_\ell\}$ be the components of~$G[V_3- v]$ with $w(C_1) \leq \ldots \leq w(C_\ell)$.}
	\If {$w(V_i) + w(v) + \sum_{j \in [\ell - 1]} w(C_j) < w(V_3)$}
		\State {\textbf{return} $U$, where $U = \{v\} \cup \bigcup_{j \in [\ell - 1]} V(C_j)$ ~~\# $U$ is pull-admissble \label{line:return:U}}
	\EndIf
        \EndFor
\State{\textbf{return} $\emptyset$}        
\EndProcedure

  
\end{algorithmic}
\caption{\textsc{PullCheck}}
\label{alg:pull}
\end{algorithm}

\begin{lemma}\label{lemma:pull-complexity}
  If there is a pull-admissible set, then Algorithm~\ref{alg:pull} finds one. Moreover, on input $(G,w)$, where $G=(V,E)$, it runs in $\bigO(|V| (|V|+|E|))$ time.
\end{lemma}
\begin{proof}
  Let $T\subset V_3$ be a pull-admissible set  w.r.t.~$i$.  We may assume that
  $G[T]$ is connected, otherwise, each of its components is
  pull-admissible and we can consider any of them.  Let~$v \in
  T$ be a vertex adjacent to~$V_i$. Clearly, $v$ will be checked at line~\ref{line:for}. Let $\mathcal{C} = \{C_1,
  \ldots, C_\ell\}$ be the components of $G[V_3 - v]$ with $w(C_1)
  \leq \ldots \leq w(C_\ell)$. (Note that~$\ell \geq 1$, as
  $T$ is a proper subset of~$V_3$.)  Since $G[V_3 \setminus T]$ and~$G[T]$ are connected, we conclude that $G[T]$ must contain $\ell - 1$ components of~$\mathcal{C}$. Moreover, precisely one of the components in~$\mathcal{C}$ is not contained in $G[T]$ (but $T$ may contain part
  of it). (To see this, consider the block structure of $G[V_3]$ and analyse when $\ell=1$ and $\ell\geq 2$.)
  It follows from this observation that the set $U= \{v\} \cup \bigcup_{j \in [\ell - 1]} V(C_j)$ is such that $w(U) \leq w(T)$. 
As $T$ is pull-admissible, it holds that~$w(V_i) + w(U) \leq w(V_i)+ w(T)  < w(V_3)$. Hence, at line~\ref{line:return:U}, the set~$U$, which is pull-admissible, will be returned by Algorithm~\ref{alg:pull}.

Since the connected components of~$G[V_3 - v]$ can be computed in
time~$\bigO(|V| + |E|)$, Algorithm~\ref{alg:pull} runs in $\bigO(|V| (|V|+ |E|))$ time.
\end{proof}



\begin{algorithm}[H]
\hspace*{\algorithmicindent} \textbf{Input:} An instance~$(G, w)$ of~$\minmaxBCP{3}$\\
\hspace*{\algorithmicindent} \textbf{Output:} A connected~$3$-partition of~$G$\\
\hspace*{\algorithmicindent} \textbf{Routines:} \textsc{Merge}, \textsc{Pull} and \textsc{PullCheck}.

\begin{algorithmic}[1]
\Procedure {\textsc{Min-Max-BCP3}}{$G, w$}

\State Let~$\mathcal{P} = \{V_1, V_2, V_3\}$ be an ordered connected $3$-partition of $G=(V,E)$;  $W=w(G)$
\While {$w(V_3) > W/2$}\label{line:loop}
	\If {$V_1$ and~$V_2$ are adjacent \textbf{and}~$|V_3| \geq 2$}
		\State {$\mathcal{P} \gets \textsc{Merge}(\mathcal{P})$} ~~\#  $\mathcal{P} = \{V_1,V_2,V_3\}$
	\ElsIf {\textsc{PullCheck}$(\mathcal{P},i)$ returns a nonempty set $U$  for $i=1$ or $i=2$}  \label{line:pull}
		\State {$\mathcal{P}\gets \textsc{Pull}(\mathcal{P}, U,i)$} ~~\#  $\mathcal{P} = \{V_1,V_2,V_3\}$
	\Else 
		\State {\textbf{break}} \label{line:break}
	\EndIf
\EndWhile

\State {\textbf{return} $\mathcal{P}$}
\EndProcedure
\end{algorithmic}
\caption{\textsc{Min-Max-BCP3}}
\label{alg:minmax3}
\end{algorithm}

\begin{lemma}\label{lemma:correctness-complexity}
  Algorithm~\ref{alg:minmax3} on input~$(G,w)$, where $G=(V,E)$ and
  $w$ is an integer-valued function, finds a connected~$3$-partition
  of~$G$ in $\bigO(w(G) |V||E|)$ time.
\end{lemma}
\begin{proof}
  The algorithm starts with an arbitrary connected $3$-partition of
  $G$, and only modifies the current partition when a \textsc{Merge}
  or \textsc{Pull} operation is performed. As both operations are performed only when the corresponding pre-conditions are satisfied, they yield connected $3$-partitions of~$G$, and the algorithm is correct.



  Each time a \textsc{Merge} or a \textsc{Pull} operation is
  executed, the weight of the heaviest class decreases. Thus, at most~$w(G)$
  calls of such operations are performed by the algorithm.  Recall
  that both \textsc{Merge} and \textsc{Pull} operations take
  $\bigO(|V|+|E|)$ time. Moreover, by  
  Lemma~\ref{lemma:pull-complexity}, the procedure \textsc{PullCheck} has time complexity~$\bigO(|V||E|)$~($G$ is connected, so~$|E| \geq |V| - 1$). It follows from the above remarks that 
  Algorithm~\ref{alg:minmax3} has time complexity  $\bigO(w(G) |V||E|)$. 
\end{proof}


It is clear that when Algorithm~\ref{alg:minmax3} halts and returns a partition~$\mathcal{P}$, one of the two cases occurs: (a) either the loop condition in line~\ref{line:loop} failed, and in this case, $\mathcal{P}$ has value $w^+(\mathcal{P}) \leq  w(G)/2$, or (b) neither \textsc{Merge} nor \textsc{Pull} operations could be performed (and $w^+(\mathcal{P}) > w(G)/2$).  
In what follows, we prove that in case~(b) the input graph has a particular ``star-like'' structure which allows us to conclude that the solution produced by the algorithm is optimal.
\begin{lemma} \label{lemma:ratio:minmax3}
Let~$\mathcal{P}=\{V_1, V_2, V_3\}$ be an ordered  connected $3$-partition produced by Algorithm~\ref{alg:minmax3}, and let $G_i=G[V_i]$, for $i=1,2,3$.
If~$|V_3| \geq 2$ and $w(V_3) > {w(G)}/{2}$,  the following hold:
\begin{enumerate}[(i)]
    \item $w(V_1) < {w(G)}/{4}$, and $V_1$ and~$V_2$ are not adjacent; and \label{item:minmax3:1}
    \item there exists~$u \in V_3$ such that~$u$ is a cut-vertex of
      $G$, $\{G_1, G_2 \} \subseteq
      \mathcal{C}$,~$w(C) \leq w(V_1)\leq w(V_2)$ for each
      $C \in \mathcal{C} \setminus\{G_1, G_2\}$, where
      $\mathcal{C}$ is the set of components of~$G-u$.  Moreover, if
      $|\mathcal{C}| = 3$ then $w(u) > w(G)/4$. \label{item:minmax3:2}
\end{enumerate}
\end{lemma}
\begin{proof}
Since~$w(V_3) > w(G)/2$, the algorithm terminated after executing line~\ref{line:break}.
This implies that neither \textsc{Merge} nor \textsc{Pull} operation can be performed on $\mathcal{P}$.
Particularly, it follows that $V_1$ and $V_2$ are not adjacent.
Moreover, note that~$w(V_1) + w(V_2) < w(G)/2$, again because~$w(V_3) > w(G)/2$.
Hence, it holds that~$w(V_1) < w(G)/4$, since $w(V_1)\leq w(V_2)$.
This proves~\ref{item:minmax3:1}.

Since $G$ is connected, and $V_1$ and $V_2$ are not adjacent, there exists $uv \in E(G)$ such that $u \in V_3$ and~$v \in V_1$.
Let~$\mathcal{C}^\prime$ be the set of components of~$G_3-u$.
Note that $\mathcal{C}^\prime \neq \emptyset$ because $|V_3|>1$, and consider a component~$C \in \mathcal{C}^\prime$.
Let us define~$S = \{u\} \cup (\bigcup_{C^\prime \in \mathcal{C} \setminus \{C\}} C^\prime )$.
It is clear that~$G[C]$ and $G[S]$ are connected subgraphs of $G$.
Since it is not possible to perform $\textsc{Pull}(\mathcal{P}, S, 1)$, it holds that $w(S) + w(C) \leq w(S) + w(V_1)$. 
Therefore, we have that $w(C) \leq w(V_1) \leq w(V_2)$.

Suppose to the contrary that that~$V_1$ is adjacent to~$C$.
Thus the partition~$\{V_1 \cup V(C), V_2, V_3 \setminus V(C) \}$ is a connected $3$-partition of $G$.
Since~$w(C) + w(V_1) \leq 2 w(V_1) < w(G)/2$, Algorithm~\ref{alg:minmax3} could perform~$\textsc{Pull}(\mathcal{P}, V(C), 1)$, a contradiction.
Similarly, $V_2$ is not adjacent to~$C$, otherwise the algorithm could execute~$\textsc{Pull}(\mathcal{P}, V(C), 2)$ because~$w(C) + w(V_2) \leq w(V_1) + w(V_2) < w(G)/2 <w(V_3)$.
By claim~\ref{item:minmax3:1}, $V_1$ and $V_2$ are not adjacent, and thus we have~$N(V_1)\cap V_3=N(V_2)\cap V_3 = \{u\}$. 
Therefore, $u$ is a cut-vertex of $G$ and~$\mathcal{C} = \{G_1, G_2\} \cup \mathcal{C}^\prime$.
This concludes the proof of~\ref{item:minmax3:2}.
\end{proof}

\begin{theorem}\label{theorem:approx:minmax3}
  Algorithm~\ref{alg:minmax3} is a pseudo-polynomial
  $\frac{3}{2}$-approximation for~$\minmaxBCP{3}$ which runs in
  $\bigO(w(G) |V| |E|)$ time on an instance $(G,w)$, where $G=(V,E)$.
\end{theorem}
\begin{proof}
Let~$\mathcal{P} = \{V_1, V_2, V_3\}$ be an ordered $3$-partition of $G$, returned by the algorithm; and let $G_i=G[V_i]$, for $i=1,2,3$.
By Lemma~\ref{lemma:correctness-complexity}, $\mathcal{P}$ is indeed a connected $3$-partition of~$G$ and it can be computed in time~$\bigO(w(G) |V| |E|)$.
If~$w(V_3) \leq w(G)/2$, then it follows directly from Fact~\ref{fact:opt} that $w^+(\mathcal{P})=w(V_3)\leq \frac{3}{2}\opt_3(G,w)$.

Suppose now that~$w(V_3) > w(G)/2$. If~$V_3$ is a singleton~$\{u\}$, then~$w(u) \leq \opt_3 (G,w)$ and~$\mathcal{P}$ is optimal. Otherwise, the algorithm
terminated because neither \textsc{Merge} nor \textsc{Pull} operation
can be performed on $\mathcal{P}$.  By
Lemma~\ref{lemma:ratio:minmax3}\ref{item:minmax3:2}, there
exists~$u \in V_3$ such that~$u$ is a cut-vertex of $G$,
$\{G_1, G_2 \} \subseteq \mathcal{C}$,
and~$w(C) \leq w(V_1)\leq w(V_2)$ for each
$C \in \mathcal{C} \setminus\{G_1, G_2\}$, where $\mathcal{C}$ is the
set of components of~$G-u$.  By Lemma~\ref{lemma:bound:max}, we have
\[w^+(\mathcal{P}) = w(V_3) = w(u)+\sum_{C\in \mathcal{C}\setminus\{G_1, G_2\}} w(C)\leq \opt_3 (G,w).\]
Therefore,  in this case the partition $\mathcal{P}$ produced by the algorithm is an optimal solution for the instance $(G,w)$ of~$\minmaxBCP{3}$.
\end{proof}

In what follows, we show how to extend the result obtained for $\minmaxBCP{3}$ to obtain results for
~$\minmaxBCP{_k}$, for all $k\geq 4$.  For simplicity, we say that a vertex~$u$ satisfying
condition~\ref{item:minmax3:2} of Lemma~\ref{lemma:ratio:minmax3} is a
\textit{star-center}. Moreover, when~$u$ is a star-center, we name the~$\ell$ components of~$G-u$
as~$\mathcal{C} = \{C_1, C_2, \ldots, C_\ell\}$, where~$C_\ell = G[V_2]$,~$C_{\ell - 1} = G[V_1]$
and~$w(C_i)\leq w(C_{i+1})$ for all~$i \in [\ell-1]$.


\begin{algorithm}[H]
  \hspace*{\algorithmicindent} \textbf{Input:} A connected graph~$G = (V, E)$, a connected~$k'$-partition~$\mathcal{P}$ of $G$, 
   and an integer~$q\geq 0$ such that $k' + q \leq |V|$ \\
\hspace*{\algorithmicindent} \textbf{Output:} A connected~$(k' + q)$-partition of~$G$

\begin{algorithmic}[1]
\Procedure {\textsc{GetSingletons}}{$G, w, q, \mathcal{P}$}
\While {$q > 0$}
	\State Let~$U \in \mathcal{P}$ be such that~$|U| \geq 2$ and~$u$ be a non-cut vertex of~$G[U]$.
	\State $\mathcal{P} \gets (\mathcal{P} \setminus \{U\}) \cup (\{\{u\}\} \cup \{U \setminus \{u\}\})$
	\State $q \gets q - 1$
\EndWhile
\State {\textbf{return} $\mathcal{P}$}
\EndProcedure
\end{algorithmic}
\caption{\textsc{GetSingletons}}
\label{alg:add-singletons}
\end{algorithm}

\begin{algorithm}[H]
\hspace*{\algorithmicindent} \textbf{Input:} An instance~$(G=(V,E), w)$ of~$\minmaxBCP{k}$, $3\leq k \leq |V|$\\
\hspace*{\algorithmicindent} \textbf{Output:} A connected~$k$-partition of~$G$ \\
\hspace*{\algorithmicindent} \textbf{Routines:} \textsc{Min-Max-BCP3},  \textsc{GetSingletons}
\begin{algorithmic}[1]
\Procedure {\textsc{Min-Max-BCP}$k$} {$G, w$}

\State $\mathcal{P} \gets \textsc{Min-Max-BCP3}(G, w)$ ~~\#  $\mathcal{P} = \{V_1,V_2,V_3\}$
\If{$w^+(\mathcal{P}) \leq \frac{w(G)}{2}$~\textbf{or}~$|V_3| = 1$}
	\State $\mathcal{P}' \gets \textsc{GetSingletons}(G, w, k - 3, \mathcal{P})$
\Else
	\State Let~$u$ be the star-center and let~$\mathcal{C} = \{C_i\}_{i\in [\ell]}$ be the components of~$G-u$.
	\If {$\ell \geq k - 1$}
		\State Let~$t = \ell - k + 1$ and~$V' = (\bigcup_{i \in [t]} V(C_i)) \cup \{u\}$.
		\State $\mathcal{P}' \gets \{V', V(C_{t + 1}), \ldots, V(C_{\ell-1}), V(C_{\ell})\}$
	\Else
		\State $\mathcal{P} \gets \{\{u\}\} \cup \{C_i\}_{i\in [\ell]}$
		\State $\mathcal{P}' \gets \textsc{GetSingletons}(G, w, k - 1 - \ell, \mathcal{P})$
	\EndIf
\EndIf
\State \textbf{return} $\mathcal{P}'$
\EndProcedure
\end{algorithmic}
\caption{\textsc{Min-Max-BCP}$k$ ~~ ($k\geq 3$)}
\label{alg:k/2:minmaxk}
\end{algorithm}

\begin{theorem}\label{theorem:approx:minmaxk}
  For each integer~$k \geq 3$, Algorithm~\ref{alg:k/2:minmaxk} is a
  pseudo-polynomial $\frac{k}{2}$-approximation for the problem $\minmaxBCPk$ that runs in
  $\bigO(w(G) |V| |E|)$ time on an instance $(G,w)$, where $G=(V,E)$.
\end{theorem}
\begin{proof}
  We first note that Algorithm~\ref{alg:add-singletons} is correct,
  since it iteratively removes~$q$ non-cut vertices from non-trivial
  classes of~$\mathcal{P}$ and create new singleton classes.
  As~$|V| \geq k' + q$, there always exists~$U \in \mathcal{P}$
  satisfying the conditions in line~3. Furthermore, note that
  if~$\mathcal{P}$ and~$\mathcal{P}'$ are the input and output of
  Algorithm~\ref{alg:add-singletons}, respectively, then it holds
  that~$w^+(\mathcal{P}') \leq w^+(\mathcal{P})$.

  Now we turn to Algorithm~\ref{alg:k/2:minmaxk}. We may assume that
  $k \geq 4$, since for~$k=3$ the result follows
  from Theorem~\ref{theorem:approx:minmax3}.  Let~$\{V_1, V_2, V_3\}$
  be the ordered connected $3$-partition produced in line~2, and let
  $G_i= G[V_i]$, for $i=1,2,3$.

  If the condition in line~3 is satisfied, then since
  Algorithm~\ref{alg:add-singletons} is correct, the partition
  $\mathcal{P}'$ (in line~4) is a connected $k$-partition of~$G$. 
  Clearly if~$V_3$ is a singleton~$\{u\}$,~$\mathcal{P}^\prime$ 
  is optimal, since~$w(u) \leq \opt_k(G,w)$. Moreover, 
  when~$w^+(\mathcal{P}) \leq w(G)/2$,
  it holds that~$w^+(\mathcal{P}^\prime) \leq w(G)/2 \leq ({k}/{2}) \opt_k(G,w)$, 
  where the last inequality is justified by Fact~\ref{fact:opt}.
  

Suppose now that~$w(V_3) > w(G)/2$ and~$|V_3| \geq 2$. 
By Lemma~\ref{lemma:ratio:minmax3}\ref{item:minmax3:2}, there exists a star-center~$u \in V_3$. Let~$\mathcal{C} = \{C_i\}_{i\in [\ell]}$ be the components of~$G-u$. We now consider two cases according to the values of $k$ and $\ell$. 

If~$\ell \geq k-1$, then in any connected $k$-partition of~$G$ the
class containing~$u$ must also contain~$t = \ell - k + 1$ components
of $G - u$. The class $V'$ defined in line~8 consists of the union of
${u}$ and the $t$~lightest such components. Clearly,
$\mathcal{P}' := \{V', V(C_{t + 1}), \ldots, V(C_{\ell-1}),
V(C_{\ell})\}$ is a connected $k$-partition of $G$, and
$w^+(\mathcal{P}') = \max\{w(V'), w(V_2)\}$ (recall that
$C_{\ell} = G[V_2]$).  If $w^+(\mathcal{P}') = w(V')$, it follows from
Lemma~\ref{lemma:bound:max} that~$\mathcal{P}'$ is an optimal
connected \hbox{$k$-partition} of~$G$.  Otherwise,
$w^+(\mathcal{P}') = w(V_2) \leq w(G)/2 \leq (k/2) \opt_k(G,w)$.


If~$\ell \leq k-2$, starting with the connected
$(\ell+1)$-partition $\mathcal{P}=\{\{u\}, V(C_1), \ldots, V(C_\ell)\}$ (as
defined in line~11), using Algorithm~\ref{alg:add-singletons} we
obtain a connected $k$-partition $\mathcal{P}'$ of~$G$.  Clearly,
$w^+(\mathcal{P}') \leq w(V_2) \leq w(G)/2$, and
so~$w^+(\mathcal{P}') \leq (k/2) \opt_k(G,w)$.

Finally, observe that non-cut vertices can be obtained by removing
leaves of any spanning tree of the graph. Thus,
Algorithm~\ref{alg:add-singletons} has time
complexity~$\bigO(|V||E|)$. Since Algorithm~\ref{alg:minmax3} (in
line~2) is a pseudo-polinomial algorithm for $\minmaxBCP{3}$ that runs
in $\bigO(w(G) |V| |E|)$ time (cf.~Theorem~\ref{theorem:approx:minmax3}), we
conclude that Algorithm~\ref{alg:k/2:minmaxk} is a pseudo-polynomial
$\frac{k}{2}$-approximation for~$\minmaxBCPk$ that runs in
$\bigO(w(G)|V||E|)$ time.
\end{proof}



The algorithm given by Theorem~\ref{theorem:approx:minmaxk} is a
(polynomial) $\frac{k}{2}$-approximation if the weights assigned to
the vertices are bounded by a polynomial on the order of the graph.
In case the weights assigned to the vertices are arbitrary, it is possible to
apply a scaling technique and use the previous algorithm as a
subroutine to obtain a polynomial algorithm for
$\minmaxBCPk$ with approximation ratio $(\frac{k}{2}+\eps)$, for any
fixed~$\eps>0$.

We prove next a more general result, concerning any pseudo-polynomial
$\alpha$-approximation algorithm for $\minmaxBCPk$ whose running time
depends on the value of the weights.


\begin{algorithm}[H]
\hspace*{\algorithmicindent} \textbf{Input:} An instance~$(G=(V,E), w)$ of~$\minmaxBCP{k}$, $3\leq k \leq |V|$\\
\hspace*{\algorithmicindent} \textbf{Output:} A connected~$k$-partition of~$G$\\
\hspace*{\algorithmicindent} \textbf{Routine:} a pseudo-polynomial $\alpha$-approximation algorithm~$\mathcal{A}$ for~$\minmaxBCP{k}$

\begin{algorithmic}[1]
\Procedure {$\eps$-\textsc{Min-Max-BCP}$k$}{$G, w$}

\State $\theta \gets \max_{v \in V} w(v)$ \label{line:theta}
\State $\lambda \gets \frac{ \eps \theta}{|V|}$ \label{line:lambda}
\For{$v \in V$}
	\State $\widehat w (v) \gets \ceil{\frac{w(v)}{\lambda}}$ \label{line:scaling}
\EndFor
\State $\mathcal{P} \gets \mathcal{A}(G, \widehat w)$ \label{line:call-approx}
\State {\textbf{return} $\mathcal{P}$}
\EndProcedure
\end{algorithmic}
\caption{$\eps$-\textsc{Min-Max-BCP}$k$~~ ($k\geq 3$)}
\label{alg:eps:minmaxk}
\end{algorithm}

\begin{theorem}\label{thm:pseudo-approximation}
  Let~$k\geq 3$ be an integer, and let $I=(G=(V,E),w)$ be an instance of $\minmaxBCPk$.
  If there is a pseudo-polynomial $\alpha$-approximation algorithm~$\mathcal{A}$ for $\minmaxBCPk$ that runs in $\bigO(w(G)^{c}|V||E|)$ time for some constant~$c$, then Algorithm~\ref{alg:eps:minmaxk} is an $\alpha(1 + \eps)$-approximation for $\minmaxBCPk$ that runs in $\bigO(|V|^{2c+1} |E|/\eps^c)$ time.
\end{theorem}
\begin{proof}

  Let $I=(G=(V,E),w)$ be an instance of~$\minmaxBCPk$, and let
  ~$\mathcal{P}^*$ (resp. $\mathcal{P}$) be an optimal solution
  (resp. a solution produced by Algorithm~\ref{alg:eps:minmaxk}) on
  input $I$. Denote by $V^*_k$ and $V_k$ the heaviest classes
  in $\mathcal{P}^*$ and $\mathcal{P}$,  respectively.
  First, note that~$\mathcal{P}^*$ is a feasible solution for the
  instance~$(G, \widehat w)$, and
  so~$\widehat w^+(\mathcal{P}^*) = \sum_{v \in V^*_k} \widehat w(v) \geq
  \opt_k(G,\widehat w)$.  Moreover,
  $\widehat w^+(\mathcal{P})=\sum_{v \in V_k} \widehat w(v) \leq \alpha
  \opt_k(G,\widehat w)$ since~$\mathcal{A}$ is an $\alpha$-approximation.
  It is clear from line~\ref{line:scaling} that $w(v)/ \lambda  \leq \widehat w(v) \leq w(v)/ \lambda +1 $ for every~$v \in V$.
  Hence, the following sequence of inequalities hold:
  \begin{align}
     w^+(\mathcal{P}) & = \sum_{v \in V_k} w(v)  \leq \lambda \sum_{v \in V_k} \widehat w(v) \leq \lambda \alpha\opt_k(G,\widehat w) \nonumber &\\
     & \leq \lambda  \alpha \sum_{v \in V^*_k} \widehat w(v) \leq \lambda  \alpha \sum_{v \in V^*_k}  \left(\frac{w(v)}{\lambda} + 1\right) & \nonumber\\
      & = \alpha \opt_k(G,w) + \lambda \alpha |V^*_k|.  &  \label{ineq:lambda}
   \end{align} 
Since $\lambda = \eps \theta / |V|$ (see line~\ref{line:lambda}) and~$\theta \leq \opt_k(G,w)$, it follows from inequality~\eqref{ineq:lambda} that 
\begin{align*}
     w^+(\mathcal{P}) & \leq \alpha \opt_k(G,w) + \alpha \eps \theta \leq  \alpha(1 + \eps)\opt_k(G,w).
\end{align*}

The running-time of Algorithm~\ref{alg:eps:minmaxk} is clearly dominated by the running-time of~$\mathcal{A}$ on input~$(G, \widehat w)$ in line~\ref{line:call-approx} which takes time~$\bigO(\widehat w(G)^c |V||E|)$.
It follows from the scaling in line~\ref{line:scaling} that $\widehat w(v) \leq w(v)/\lambda +1 \leq |V|/\eps +1$ for every~$v \in V$.
Therefore, $\widehat w(G) \leq |V|^2/\eps + |V|$, and thus, the algorithm runs in $\bigO(|V|^{2c+1}|E|/\eps^c)$ time.
\end{proof}

\bigskip

\begin{corollary}
  For each integer~$k \geq 3$ and $\eps'>0$, there is
  a~$(\frac{k}{2}+ \eps')$-approximation for~$\minmaxBCPk$ that runs in
  $\bigO(|V|^3 |E|/\eps')$ time on a input $(G=(V,E),w)$. 
\end{corollary}

\begin{proof}
  The result follows from Theorem~\ref{thm:pseudo-approximation}, by
  taking Algorithm~\ref{alg:eps:minmaxk} with $\eps = \eps'/(k/2)$ and
  Algorithm~\ref{alg:k/2:minmaxk} as the routine~$\mathcal{A}$ it
  requires. The approximation ratio~$k/2$ of
  Algorithm~\ref{alg:k/2:minmaxk} is guaranteed by
  Theorem~\ref{theorem:approx:minmaxk}.
\end{proof}

\bigskip

An algorithm analogous to Algorithm~\ref{alg:eps:minmaxk} can be designed for $\maxminBCPk$. In this case, 
change line~\ref{line:theta} to  $\theta \gets \min_{v \in V} w(v)$, change line~\ref{line:scaling} to  $\widehat w (v) \gets \floor{\frac{w(v)}{\lambda}}$, and consider a routine that is a pseudo-polynomial $\alpha$-approximation for $\maxminBCPk$. Then, a  theorem similar to Theorem~\ref{thm:pseudo-approximation} can be obtained for  $\maxminBCPk$.

\section{Parameterized $\maxminBCPsemk$}



This section is devoted to design a fixed-parameter tractable (FPT) algorithm based on integer linear programming for the max-min version of the unweighted balanced connected partition problem when parameterized by the vertex cover.
In this problem, we are given an unweighted graph~$G$, a positive integer~$k$, and a vertex cover~$X$ of~$G$.
The objective is to find a connected $k$-partition of~$G$ that maximizes the size of the smallest class.
Let us consider a fixed instance~$(G,k)$ of \textsc{max-min-BCP} and a vertex cover~$X$ of~$G$.

Let us denote by~$I$ the stable set~$V(G)\setminus X$.
Recall that we assume~$k \leq |V(G)| = |X|+|I|$.
If~$k > |X|$, then there are at least $k-|X|$ classes of size exactly~1 contained in~$I$, and so an optimal solution (which has value equal to~1) can be easily computed.
If~$|X|=1$, then~$G$ is a star, and so it is trivial to compute an optimal solution.
From now on, we assume that~$k\leq|X|$ and $|X|\geq 2$.

Before presenting the details of the proposed algorithm, we prove a
simple lemma that guarantees the existence of an optimal solution in
which each class intersects the given  vertex cover~$X$.

\begin{lemma}\label{lemma:X:solution}
  Let $(G,k)$ be an instance of \textsc{max-min-BCP} and let~$X$ be a vertex cover of~$G$.
  Then, there exists an optimal connected $k$-partition~$\{V_i\}_{i\in [k]}$ of~$G$ such that $V_i \cap X \neq \emptyset$ for all~$i \in [k]$.
\end{lemma}
\begin{proof}
	Suppose to the contrary that no such partition exists.
	Let~$\{V^\prime_i\}_{i\in [k]}$ be a connected $k$-partition of~$G$ with the smallest number of classes contained in~$I$, and let~$V^\prime_j = \{v\} \subseteq I$, for some~$j \in [k]$, be one of these classes. 
	Since~$k\leq |X|$, there exists  $\ell \in [k]\setminus\{j\}$ such that~$|V^\prime_\ell \cap X|\geq 2$.
	One may easily find a partition~$\{V^\prime_{\ell,1}, V^\prime_{\ell,2}\}$ of~$V^\prime_\ell$ such that, for~$i \in \{1,2\}$, $G[V^\prime_{\ell,i}]$ is connected and~$V^\prime_{\ell, i} \cap X \neq \emptyset$.
	If $N(v)\cap V^\prime_\ell \neq \emptyset$, then assume without loss of generality that $N(v)\cap V^\prime_{\ell,1} \neq \emptyset$.
	In this case, there is a connected $k$-partition~$\{V_i\}_{i\in[k]}$ of $G$ such that $V_j = V^\prime_{\ell,1} \cup \{v\}$, $V_\ell = V^\prime_{\ell,2}$ and $V_i = V^\prime_i$ for every~$i\in [k]\setminus\{j,\ell\}$.
	If $N(v)\cap V^\prime_\ell = \emptyset$, then there exists~$t \in [k]\setminus\{j, \ell\}$ such that $N(v)\cap V^\prime_t \neq \emptyset$ since $G$ is connected.
	Clearly, such a class intersects $X$, that is, $V^\prime_t\cap X \neq \emptyset$.
	Thus, there is a connected $k$-partition~$\{V_i\}_{i\in[k]}$ of $G$ such that $V_j = V^\prime_{\ell,1}$, $V_\ell = V^\prime_{\ell,2}$, $V_t=V^\prime_t \cup \{v\}$ and $V_i = V^\prime_i$ for every~$i\in [k]\setminus\{j,\ell, t\}$.
	In both cases, the partition has a smaller number of classes contained in $I$ than~$\{V^\prime_i\}_{i\in[k]}$, a contradiction to the choice of this partition.
\end{proof}

We next use hypergraphs to model the constraints of our formulation for~\textsc{max-min-BCP}.
A hyperpath of length~$m$ between  two vertices~$u$ and~$v$ in a hypergraph~$H$ is a set of hyperedges~$\{e_1, \ldots, e_m\} \subseteq E(H)$ such that~$u \in e_1$, $v \in e_m$, and $e_i\cap e_{i+1} \neq \emptyset$ for each~$i  \in \{1,\ldots, m-1\}$.
A set of hyperedges~$F \subseteq E(H)$ is a $(u,v)$-cut if there is no hyperpath between~$u$ and~$v$ in $H-F$.

For each~$S\subseteq X$, we define~$I(S) = \{ v \in I \colon N(v)=S\}$.
Let~$u,v \in X$ be a pair of non-adjacent vertices in $G$, and let~$\Gamma_X(u,v)$ be the set of all separators of $u$ and~$v$ in~$G[X]$.
Consider a separator~$Z \in \Gamma_X(u,v)$, and denote by~$\mathcal{C}(Z)$ the set of components of~$G[X-Z]$.
Let~$H_Z$ denote the hypergraph with vertices~$\mathcal{C}(Z)$ such that, for each~$S \subseteq X$ with $I(S)\neq \emptyset$, there is a hyperedge~$\{C \in \mathcal{C}(Z) \colon S\cap V(C) \neq \emptyset\}$ in~$H_Z$.
We denote by~$\Lambda_Z(u,v)$ the set of all $(C_u,C_v)$-cuts in~$H_Z$, where $C_u$ and $C_v$ are the components of $G[X-Z]$ containing $u$ and~$v$, respectively.


For each~$v \in X$ and $i  \in [k]$, there is a binary variable~$x_{v,i}$ that equals~1 if and only if~$v$ belongs to the $i$-th class of the partition.
Moreover, for every~$S \subseteq X$ and $i \in [k]$, there is an integer variable~$y_{S,i}$ that equals the amount of vertices in~$I(S)$ that are assigned to the $i$-th class.
The intuition for the meaning of the $y$-variables is that all vertices in~$I(S)$, for a fixed~$S\subseteq X$, play essentially the same role in a connected partition. 
Hence, the formulation does not need to have decision variables associated with the vertices in~$I(S)$, and so it only has an integer variable to count the number of these vertices that are chosen to be in each class.
The idea of using integer variables to count indistinguishable vertices in a stable set was used before by Fellows et al.~\cite{fellows2008graph} for the \textsc{Imbalance} problem.

Let us define~$\eta=2^{|X|}$, that is, $\eta$ is the number of subsets of $X$.
Let~$\mathcal{B}(G,X,k)$ be the set of vectors in $\R^{(|X|+\eta)k}$ that satisfy the following inequalities~\eqref{ineq:asymetric}--\eqref{ineq:int}.

\begin{align} 
    &\sum_{v \in X} x_{v,i} + \sum_{S\subseteq X} y_{S,i}  \leq \sum_{v \in X} x_{v,i+1} + \sum_{S\subseteq X} y_{S,i+1} &  \!\!\!\!  \forall i \in [k-1],\label{ineq:asymetric} &\\ 
    &\sum_{i \in [k]} x_{v,i}  = 1 &   \!\!\!\! \forall v \in X,\label{ineq:cover} &\\ 
    &x_{u,i} +  x_{v,i} - \sum_{z \in Z} x_{z,i} - \sum_{S \in F} y_{S,i} \leq 1 &   \!\!\!\! \forall  uv \notin E(G[X]), Z \in \Gamma_X(u,v), F \in \Lambda_Z(u,v), i \in [k], \label{ineq:cut} &\\
    & y_{S,i} \leq |I(S)|\left(\sum_{v \in S}x_{v,i}\right) &   \forall  S \subseteq X, i\in[k],  \label{ineq:neighborX}&\\
    &\sum_{i\in[k]} y_{S,i} = |I(S)| &   \!\!\!\! \forall  S \subseteq X,\label{ineq:S:size}&\\ 
   & x_{v,i} \in \{0,1\} &  \!\!\!\! \forall v \in X \text{ and } i \in [k],\label{ineq:bin}&\\
    &y_{S,i} \in \Z_\geq &  \!\!\!\! \forall S \subseteq X \text{ and } i \in [k].\label{ineq:int}&
\end{align}
\smallskip

Inequalities~\eqref{ineq:asymetric} establish a non-decreasing ordering of the classes according to their sizes.
Inequalities~\eqref{ineq:cover} and~\eqref{ineq:S:size} guarantee that every vertex of the graph belongs to exactly one class (i.e. the classes define  a partition).
Due do Lemma~\ref{lemma:X:solution}, we may consider only partitions such that each of its classes intersects~$X$.
Thus, whenever a vertex in the stable set~$I$ is chosen to belong to some class, at least one of its neighbors in~$X$ has to be in the same class.
This explains the meaning of inequalities~\eqref{ineq:neighborX}.
Inequalities~\eqref{ineq:cut} guarantee that each class of the partition induces a connected subgraph.
This will be more clear in the proof of the following proposition.

\begin{lemma}\label{lem:BCP_ILP}
	Let~$G$ be a connected graph, let $k\geq 2$ be an integer, and let~$X$ be a vertex cover of~$G$.
	The problem \textsc{max-min-BCP} on instance~$(G,k)$ is equivalent to
	\[\max \left\{\sum_{v \in X} x_{v,1} + \sum_{S\subseteq X} y_{S,1} \colon (x,y) \in \mathcal{B}(G,X,k)\right\}.\]
\end{lemma}
\begin{proof}

	Let~$\{V_i\}_{i \in [k]}$ be a connected $k$-partition  of $G$ such that~$V_i\cap X \neq \emptyset$ for every~$i \in [k]$.
	Suppose further that the classes in this partition are ordered so that $|V_i| \leq |V_{i+1}|$ for all $i\in [k-1]$.
	From~$\{V_i\}_{i \in [k]}$, we construct a vector~$(\bar x, \bar y ) \in \R^{|X|k} \times \R^{\eta k}$ such that its non-null entries are precisely defined as follows.
	For each~$i \in [k]$, we set~$\bar x_{v,i}=1$ for every $v \in X \cap V_i$, and $\bar y_{S,i} = |I(S)\cap V_i|$ for every~$S \subseteq X$.

	We next show that $(\bar x, \bar y)$ satisfies inequalities~\eqref{ineq:asymetric}--\eqref{ineq:int}.
	It easily follows from the construction of the vector and from the hypothesis on~$\{V_i\}_{i \in [k]}$ that inequalities~\eqref{ineq:asymetric},\eqref{ineq:cover},\eqref{ineq:S:size},\eqref{ineq:bin}, and~\eqref{ineq:int} hold for~$(\bar x, \bar y)$.
	Moreover, since $V_i\cap X \neq \emptyset$ for every~$i \in [k]$, inequalities~\eqref{ineq:neighborX} hold for~$(\bar x, \bar y)$.
	It remains to prove that inequalities~\eqref{ineq:cut} are satisfied.
	
	Consider an integer~$i \in [k]$ such that there is a pair of non-adjacent vertices $u,v \in X\cap V_i$.
	Let~$Z \subseteq X\setminus\{u,v\}$  be a separator of~$u$ and~$v$ in~$G[X]$.
	Since $G[V_i]$ is connected, there exists a path~$P$ in this graph with endpoints~$u$ and~$v$ such that either~$V(P) \cap Z \neq \emptyset$ or $V(P)\cap I \neq \emptyset$. 
	Suppose that~$V(P) \cap Z = \emptyset$, otherwise inequalities~\eqref{ineq:cut} for~$Z$ are clearly satisfied by~$(\bar x, \bar y)$.
	Hence there is a hyperpath in~$H_Z$ linking $C_u$ and $C_v$, where $C_u$ and~$C_v$ are the components of~$G[X-Z]$ containing $u$ and $v$, respectively.
	As a consequence, for each cut~$F$ separating~$C_u$ and~$C_v$ in the hypergraph~$H_Z$, there exists a vertex~$z \in V(P)\cap I$ such that $N(z) \in F$, and so~$\bar y_{N(z), i} \geq 1$.
	Therefore, inequalities~\eqref{ineq:cut} are satisfied by~$(\bar x, \bar y)$.

	Let~$(\bar x, \bar y) \in \mathcal{B}(G,X,k)$. 
	Consider a subset~$S \subseteq X$.
	It follows from inequality~\eqref{ineq:S:size} for~$S$ that there is a partition~$\{I_i(S)\}_{i \in [k]}$ of~$I(S)$ such that $|I_i(S)| = \bar y_{S,i}$.
	We remark that some classes in this partition may be empty.
	For each~$i \in [k]$, let us define~$V_i =(\bigcup_{S\subseteq X}I_i(S)) \cup \{v \in X \colon \bar x_{v,i}=1\}$.
	One may easily verify that~$|V_i| = \sum_{v \in X} \bar x_{v,i} + \sum_{S\subseteq X} \bar y_{S,i}$.
	Observe now that~$I(S)\cap I(S^\prime)  = \emptyset$ for all~$S, S^\prime$ subsets of $X$ with $S\neq S^\prime$, and thus  $\{I(S)\}_{S\subseteq X}$ is a partition of~$I$ with possibly some empty classes.
	Furthermore, inequalities~\eqref{ineq:cover}  guarantee that each vertex in~$X$ belongs to exactly one class $V_i$, for some $i \in [k]$.
	Therefore, $\{V_i\}_{i\in [k]}$ is a partition of the vertices in~$G$.
	Due to  inequalities~\eqref{ineq:asymetric}, it also holds that~$|V_i| \leq |V_{i+1}|$ for all~$i \in [k-1]$. 
	We shall prove that $G[V_i]$ is connected for each~$i  \in [k]$.

	Suppose to the contrary there exists~$i \in [k]$ such that~$G[V_i]$ is not connected.
	Because of  inequalities~\eqref{ineq:neighborX},  every component of~$G[V_i]$ has to intersect~$X$.
	Let us define~$Z= X \setminus V_i$, and let~$H_Z$ be the hypergraph of the components of $G[X-Z]$ as defined earlier.
	It follows that, for each hyperedge~$S$ of $H_Z$, no vertex in~$I(S)$ belongs to~$V_i$.
	Hence, for every pair of vertices~$u, v \in V_i\cap X$ belonging to distinct components of~$G[V_i]$,  it holds that 
	\[\bar x_{u,i} + \bar x_{v,i} - \sum_{z \in Z} \bar x_{z,i} - \sum_{S \in E(H_Z)} \bar y_{S,i} = 2 > 1.\]
	This is a contradiction to the fact that $(\bar x, \bar y)$ satisfies inequalities~\eqref{ineq:cut}.
	As a consequence, we conclude that~$G[V_i]$ is connected for each~$i \in [k]$.
	Therefore~$\{V_i\}_{i\in [k]}$ is a connected $k$-partition of $G$.

	Finally, it follows from Lemma~\ref{lemma:X:solution} that the proposed integer linear program has an optimal solution that corresponds to an optimal connected $k$-partition of~$G$. 
	As a consequence, it is equivalent to solving the \textsc{max-min-BCP} problem on instance~$(G,k)$.
\end{proof}

The main tool to design fixed-parameter tractable algorithms using integer linear programing (ILP) is a theorem due to Lenstra~\cite{Len83} which shows that checking the feasibility of an ILP problem  with a fixed number of variables can be solved in polynomial time.
The time and space complexity of Lenstra's algorithm were later improved by~Kannan~\cite{Kan87}, and~Frank and Tardos~\cite{FraTar87}.
In this work, we consider the following optimization version of their results.

In the \textsc{Integer Linear Programming} problem, we are given as input a matrix~$A \in \Z^{p\times q}$, vectors~$b \in \Z^p$ and~$c \in \Z^q$.
The objective is to find a vector~$x\in \Z^q$ that satisfies all inequalities (i.e. $Ax\leq b$), and maximizes~$c^T x$.
Let us denote by~$L$ the size of the binary representation of an input~$(A,b,c)$ of the problem.


We next present the maximization version of the theorem showed in Cygan et al.~\cite{Cygan2015ChapterILP} on the existence of an FPT algorithm for \textsc{Integer Linear Programming} parameterized by the number of variables.

\begin{theorem}[Cygan et al.~\cite{Cygan2015ChapterILP}]\label{thm:FPT_ILP}
	An \textsc{Integer Linear Programming} instance of size~$L$ with~$q$ variables can be solved using
	$\bigO(q^{2.5q + o(q)} \cdot (L + \log M_x) \log(M_x M_c) )$
arithmetic operations and space polynomial in $L + \log M_x$, where $M_x$ is an upper bound on the absolute value a variable can take in a solution, and $M_c$ is the largest absolute value of a coefficient in the vector~$c$.
\end{theorem}

The previous theorem is now used to show that the max-min unweighted version of the balanced connected partition problem admits an algorithm that runs in time doubly exponential in the size of a vertex cover of the input graph.

\begin{theorem}
	The problem \textsc{max-min-BCP}, parameterized by the size of a vertex cover of the input graph, is fixed-parameter tractable.
\end{theorem}
\begin{proof}
	Consider an instance~$(G,k)$ of \textsc{max-min-BCP}, and a vertex cover~$X$ of~$G$.
	It follows from Lemma~\ref{lem:BCP_ILP} that $\max \left\{\sum_{v \in X} x_{v,1} + \sum_{S\subseteq X} y_{S,1} \colon (x,y) \in \mathcal{B}(G,X,k)\right\}$ is equivalent to solving instance~$(G,k)$.
	Observe now that the size of this integer linear program (ILP) is $2^{2^{\bigO(|X|)}} \log |G|$.
	By Theorem~\ref{thm:FPT_ILP},  this ILP problem can be solved in time~$2^{2^{\bigO(|X|)}} |G|^{\bigO(1)}$.
	Therefore \textsc{max-min-BCP} is fixed parameter-tractable when parameterized by the size of a vertex cover of the input graph.
\end{proof}


\bibliographystyle{abbrv}
\bibliography{BCP-bibliography}

\end{document}